\documentclass[11pt,twoside]{article}

\usepackage{asp2014}

\aspSuppressVolSlug
\resetcounters

\begin{document}

\title{TOPCAT: Working with Data and Working with Users}

\author{M.~B.~Taylor
  \affil{H.~H.~Wills Physics Laboratory, University of Bristol, UK;
         \email{m.b.taylor@bristol.ac.uk}}
}

\paperauthor{M.~B.~Taylor}{m.b.taylor@bristol.ac.uk}{0000-0002-4209-1479}{University of Bristol}{School of Physics}{Bristol}{Bristol}{BS8 1TL}{U.K.}

\begin{abstract}
TOPCAT is a desktop application for interactive analysis of tabular data,
especially source catalogues.  Along with its command-line counterpart
STILTS, it has been under more or less continuous development for
the past 15 years and is now widely used by astronomers from
project students to research scientists.
This paper reviews its capabilities
as a tool for working with large and small datasets,
and considers some of the issues in design, implementation and
user interaction that have to be tackled when developing software
of this kind.
\end{abstract}


\section{Introduction}

TOPCAT,\footnote{\url{http://www.starlink.ac.uk/topcat/}}
the Tool for OPerations on Catalogues and Tables,
is a desktop interactive data analysis application
for working with tabular data.
It is typically used with astronomical source
catalogues, but it can also be applied to other kinds of table
within or outside of astronomy.
The purpose of the software is to take care of all the mechanical
operations that astronomers need to perform when working with tables,
so that they can concentrate on understanding the scientific
meaning buried in the data.

The first release of TOPCAT, in 2003,
was as a fairly simple viewer and calculator for data and metadata
of tables with I/O support for a number of file formats,
including the then-new VOTable.
It has been under more or less continuous development since then,
and now provides many features including column calculations,
graphical or programmatic row selections,
highly configurable and scalable visualisation,
flexible crossmatching, access to Virtual Observatory services,
and more.
Since 2005 it has been accompanied by a sister package
STILTS,\footnote{\url{http://www.starlink.ac.uk/stilts/}}
which provides a command-line interface to the same functionality.
In 2017, it is an established part of the astronomy software landscape,
with an active user base of a few thousands
from undergraduates to research scientists
(along with a few amateur astronomers, high school students and
enthusiasts from outside astronomy),
and a few hundred citations in the literature
\citep{2005ASPC..347...29T}.
Essentially all of the development has been done by the current author.

This paper discusses some of the problems that the author
has had to address in the course of this ongoing development,
along with the approaches taken to solve them,
with varying degrees of success.
This discussion is by no means intended as the definitive manual on how
to develop application software, but it is presented as a case study
in the hope that it may be illuminating to other individuals or groups
with similar software development aims.

The problems encountered may be divided into two categories:
those that are mainly technical in nature
such as choice of deployment platform or
implementation strategies for efficient data access,
and those centred on more human considerations,
such as user engagement and user interface design.
Those in the former category tend to admit of fairly
well-defined (though not necessarily unique) solutions,
while the latter are more open-ended and more challenging.
For this reason
it can be tempting to focus on the technical issues at the
expense of the human ones.
This however risks producing software that is technically
excellent but not much used,
so for software that is successful in the sense of wide take-up,
it is essential to pay attention to both classes of problem.

The paper is organised as follows.
Section~\ref{sec:I3-1_review} reviews the
capabilities of the TOPCAT application for context.
Section~\ref{sec:I3-1_tech} discusses some of the technical questions
encountered during development
along with the solutions adopted to address them.
Section~\ref{sec:I3-1_human} presents a selection of
issues more influenced by human factors;
for these less tractable problems
no pretence is made of providing solutions,
but some best guess approaches and guidelines are discussed.

\section{TOPCAT Overview}
\label{sec:I3-1_review}

TOPCAT is a data analysis application.
Its job is to give astronomers the tools they need to
extract maximum value from one type of data product, tables,
especially those
produced by the complex, beautiful and expensive detectors
built to survey the skies.

General requirements for a data analysis application include
capabilities for data acquisition, metadata inspection,
data representation, data integration, data manipulation,
and ideally the possibility to combine both exploratory and
reproducible modes of control.
This section presents brief discussions of these requirements,
along with the features that TOPCAT/STILTS provides to satisfy them,
thus giving an overview of its main capabilities.

{\bf Data Acquisition:}
TOPCAT offers various ways to import tables into the application.
The most straightforward is to load data directly from an existing
local file.  Various formats are supported, including
FITS, VOTable, CDF and a modest selection of text-based formats
including CSV.
Additionally, quite extensive capabilities for external data access
are provided, in particular to services implementing a number of
standardised {\em Virtual Observatory\/} protocols.

{\bf Metadata Inspection:}
The availability and importance of metadata is dependent on its source.
For a table loaded, e.g., from a local CSV file, available metadata will be
limited to column names at most.
Where the user is familiar with the data and its provenance
this may be quite adequate.
But for complex datasets acquired from remote services whose content
and processing are not well understood by the user,
per-column metadata such as units and textual descriptions,
and per-table items such as coordinate system information or
query execution parameters, may be essential to understand
the attached data.
TOPCAT provides GUI components for displaying and editing such
metadata, and attempts to preserve it over save/load cycles
provided a suitably capable serialization format is used.

{\bf Data Representation:}
TOPCAT provides a data examination window which allows the user to
browse cell data of loaded tables.
In most cases however, the volume of data makes this of limited use,
so a wide range of visualisation capabilities is offered,
allowing the user to plot columns against each other
using many variations
on the theme of a point cloud in one, two or three spatial dimensions,
with higher dimensionality representable using attributes such as
colour, shape, orientation, labelling etc,
and further explorable using {\em linked views}
\citep{Tukey77,2012AN....333..505G}.
Special attention is given to representing large datasets in ways
that are meaningful (consider for example the problem of
representing $10^7$ points on a $10^6$-pixel grid)
as well as performant,
though tables of tens or hundreds of points are equally well served.
The visualisation capabilities are perhaps the most prominent of
TOPCAT's features; this functionality and its implementation are
discussed in more detail elsewhere
\citep{2014ASPC..485..257T,informatics4030018}.

{\bf Data Integration:}
In an era of multi-wavelength and multi-messenger astronomy,
much scientific output comes from a synoptic view of multiple
datasets (for instance, multiple observations of the same population
at different wavelengths) rather than simply examining a single
catalogue.
TOPCAT provides a flexible range of crossmatching options
to integrate data from different input sources;
matching two tables by sky position with fixed or per-object errors
is the most common requirement,
but other options such as matching within a single table or between
three or more tables, and on criteria such as an $N$-dimensional
Cartesian position or a unique identifier are also available.
Some of this functionality is performed internally,
and some (especially where at least one of the tables is too
large for local download) interfaces with external services
such as the CDS X-Match service \citep{2011ASPC..442...85P},
upload-capable TAP servers, or by issuing multiple Cone Search requests.

{\bf Data Manipulation:}
TOPCAT offers various options to add, delete, or reorder table columns,
identify row selections, and perform column calculations.
This is partly based on the third-party Java Expressions Library (JEL)
which allows the user to write expressions using column names as
variables in a familiar but powerful syntax,
with an extensive and extensible library of generic
(e.g. trigonometry, string manipulation, conditional evaluation)
and astronomy-specific
(e.g. flux/magnitude conversion, sky distance calculation,
time format manipulation)
functions.
The resulting expressions can define new columns,
be plotted directly, or specify selection criteria.

{\bf Exploratory and Reproducible Control:}
TOPCAT's graphical user interface lends itself well to
interactive exploration of a dataset to uncover suspected
or unexpected features.
The various capabilities for inspection of the data and metadata
described above provide a powerful platform from which to
investigate data whose form or content is not initially well
understood by the user, in order to extract scientific meaning.
However, in some cases the basic form of a dataset is
already understood, and some well-defined sequence of
operations needs to be carried out on it.
For these cases TOPCAT's point-and-click interface is less suitable,
so the sister package STILTS \citep{2006ASPC..351..666T} provides a
command-line interface to all the same functionality.
The learning curve for STILTS is somewhat steeper than for TOPCAT,
especially for visualisation operations that can require complex
specifications, and has to date been much less widely used.
To help address this, a capability has been added in the
most recent TOPCAT release (v4.5)
that allows users to set up visualisations in TOPCAT and
easily export the STILTS command that would generate
the same image.

{\bf Scalability:}
TOPCAT is not generally intended to work directly with the
largest current survey catalogues,
but it can deal with fairly large datasets.
Tables with the order of $10^6$ rows are easily handled in TOPCAT
on even low-end computers; $10^7$ is generally feasible though
with reduced responsiveness, and some users report usage with
$10^8$ rows or more.  Resource usage is less sensitive to column count;
a few hundred columns presents no problem.
STILTS has somewhat different implementation constraints,
and can for many operations stream input data to work with
tables of unlimited size in fixed memory.

\section{Implementation Notes}
\label{sec:I3-1_tech}

This section discusses some of the technical questions encountered
and solutions chosen during development,
implementation and deployment of the TOPCAT software.


\subsection{External Libraries}

TOPCAT makes use of a number of external open-source libraries.
Notable libraries include the Java 2 Standard Edition itself
for basic utilities, GUI components, XML processing, system interaction etc;
the Java Expressions Library for evaluation of user-entered expresssions;
the PixTools HEALPix library,
the CDS/ARI ADQL library, the CDS MOC library, and others.
These are used where the required functionality is clearly distinct
from the rest of the application and the implementations appear to
be robust and trustworthy.

Various other functions for which external libraries exist, however,
have been implemented using custom code.
In the case for instance of 
FITS bulk I/O, plot rendering and command-line handling,
the behaviour is so integral to the application,
in terms of functionality and performance characteristics required,
that it is both cheaper in development effort,
and provides superior capabilities,
to implement them from scratch
than to integrate with or adapt off-the-shelf libraries.
In other cases, available libraries provide much more general
capabilities than are required, and ingesting them into the
application could significantly complicate the
build process or enlarge the distributed binary.
For instance when a bare-bones internal HTTP server was required,
it was easier to implement one from scratch (approx 36\,kbyte of
compiled classes) than to link to the then-available
Jetty (1400\,kbyte).\footnote{I am indebted to Pierre Fernique (CDS)
for pointing this out to me.}

To summarise, external libraries are an essential part of most
software development, but they should be selected carefully,
and sometimes reinventing the wheel can be the better option.

\subsection{External Software}

TOPCAT is sometimes described as a Virtual Observatory (VO) application.
While many of its functions are
independent of network activity,
it does derive much of its usefulness from
interfacing with VO services, such as
Cone Search \citep{2008ivoa.specQ0222P},
TAP \citep{2014A&C.....7...37N}, 
and the VO Registry
\citep{2015A&C....10...88D}.
The great advantage of these services derives from the fact that
multiple data providers can present their diverse data holdings
using a single standard interface.
TOPCAT can therefore provide a single client per protocol
which is able to access data from many different VO-compliant archives.
This benefits the application developer,
since client code need only be written once,
and also the user, since only one user interface needs to be learnt.
These benefits are made possible by the considerable effort expended
by the working groups of the International Virtual Observatory Alliance
(IVOA) over the last decade or so in defining these data access
protocols and the many related standards on which they build.

TOPCAT interfaces with a few non-standard services as well,
in cases where some functionality is uniquely provided from a
single source, for instance VizieR, the CDS X-Match service,
and ARI's Global TAP Schema.

It also makes use of the Simple Application Messaging Protocol, SAMP
\citep{2015A&C....11...81T}.
Although TOPCAT's target is tables, comparison with other
data products such as images or spectra may be required.
To avoid having to implement for instance image-related display
and manipulation internally, it instead communicates with the
loosely-integrated suite of SAMP-aware desktop tools,
so that image-related functions can be left instead to
dedicated image analysis tools such as
Aladin \citep{2000A&AS..143...33B} or
SAOImage ds9 \citep{2003ASPC..295..489J}.

\subsection{Data Access Model}

TOPCAT uses a straightforward model for data access:
the user identifies and retrieves to local storage a table
of interest and then performs operations on it.
Such tables may be acquired from various different sources,
but once loaded there is no continuing connection with a
data server, and if more rows are required the user has to
initiate another load operation.
This is in contrast to more sophisticated data access models
such as that offered by the HiPS system \citep{2015A&A...578A.114F},
which performs loading on demand of an appropriate level of detail
from a large hierarchically prepared dataset held on a remote server.

The two approaches have different pros and cons.
TOPCAT's low-technology approach is not suitable for working
directly with the largest surveys, but a subset of manageable
size can usually be identified for download,
and it has the advantage that no advance preparation,
prior assumptions about hierarchical organisation,
or special server-side support is required.
It is also robust against connectivity issues;
once a table has been acquired and saved to local disk,
it can be used without a network connection,
or if the service that supplied it becomes temporarily unavailable
or permanently retired.

\subsection{Local I/O}

In view of the central use of client-side saved data
described in the previous section, efficient access to the data of large
tables on local disk is crucial for good performance
of the application.
TOPCAT makes use of the FITS binary table format
\citep{2010A&A...524A..42P} for caching and persisting tables
to local storage.  This format provides predictable layout of
rows and columns, so that having once read the table header,
software can instantly determine the file offset for
any given table cell.
This predictable layout can be combined with the
{\em memory mapping\/} operation offered by most operating systems,
so that each cell of a table can be addressed as if it resided
in a normal memory location rather than requiring explicit read
operations.
This approach works even if the table is much larger than physical RAM,
and benefits from optimised page caching with little
effort from application or library code, since the I/O management
is handled transparently by the OS.
In practice it delivers very efficent sequential and random access,
along with effectively instant loading, for even very large tables
\citep{2008ASPC..394..422T}.

While FITS is good for bulk data storage, its capabilities for
metadata storage are fairly primitive.
TOPCAT therefore arranges to store rich metadata,
in a serialization borrowed from the VOTable format,
in otherwise unused parts of the FITS file.
It also uses one or two other private conventions,
to provide column-oriented storage in some cases and
to work round the FITS limitation of 999 table columns
if required.
The files it writes in this way are legal FITS, but in some cases
other FITS readers may see less content than TOPCAT is
able to.

\subsection{Scalability}

TOPCAT, and especially the library code underlying it,
takes great care to impose as few restrictions as possible
on the size of dataset on which it can operate.
This kind of scalability is not easy to retrofit onto
existing software, it has to be built in from the ground up.

To achieve this, when processing potentially large collections,
generally of table rows, care is taken to avoid assumptions that
might be violated if the collection is large or even unbounded (streamed).
The following techniques are used in the code:
\begin{itemize}
\vspace*{-1.5ex}
\item Data access is performed via polymorphic abstractions (java interfaces)
      rather than assuming any particular data structure
      (such as a java array).
\vspace*{-1.5ex}
\item 64-bit rather than 32-bit integers are used to index
      or count collections.
\vspace*{-1.5ex}
\item Algorithms using indexed (random) access are avoided in
      favour of ones using (sequential) iterators.
\vspace*{-1.5ex}
\item Algorithms requiring memory that scales with collection size
      are avoided if at all possible.
\vspace*{-1.5ex}
\end{itemize}
A case in point for the last item is for visualisation.
Most off-the-shelf interactive plotting libraries allocate an object
or a few bytes for each plotted point.
Most of the plotting routines in TOPCAT instead
allocate a fixed grid of pixels and populate it progressively
while iterating over each plotted point.
This enables generation of scatter plots or density maps of arbitrarily
large tables in fixed memory.


\subsection{Deployment Platform}

TOPCAT is implemented in pure Java.
The Java language has many features which have facilitated development,
including a solid base of library classes, static typing,
and good concurrency support amongst other things.
Probably the most beneficial factor however
has been the insulation from installation environment
ensured by its Virtual Machine-based architecture.
Software build is done once, and distribution is essentially a
case of providing a single architecture-independent ``jar'' file,
avoiding the necessity to build on or support
multiple OSes or OS versions,
and facilitating installation from the user point of view.

The restriction to pure Java brings disadvantages as well.
One is that C-based libraries are essentially inaccessible;
for this reason TOPCAT has for instance no HDF5 support,
and integration with CPython is not straightforward.
Integration with OS-specific desktop features can also be
imperfect.

A current trend in interactive astronomy software,
for instance
Aladin Lite \citep{2014ASPC..485..277B} and
Firefly \citep{2013ASPC..475..315R},
is migration into the browser,
and the question is sometimes raised of whether TOPCAT's functionality
will or should be made available as a web application.
Such a move would certainly present some advantages,
not least reducing the barrier to entry for beginning users.
However, the sandboxing imposed by browsers precludes
the memory-mapped access to bulk client-side data that underlies
TOPCAT's performance with large tables.
Considering as well issues relating to multi-environment support,
the difficulty of presenting the GUI in a single browser window,
and the implementation effort required,
a wholesale port of TOPCAT to a web application is not
under consideration for the forseeable future.
Some use of the library code to provide
server-side visualisation with an interactive browser-based front end
may however be investigated.

\section{User Considerations}
\label{sec:I3-1_human}

This section presents a selection of issues concerned more with
human psychology than with the manipulation of bytes.
Some pointers and guidelines to addressing them are given,
but in most cases this paper does not have straightforward
solutions to present.

\subsection{Take Up}

One criterion of software success
is how widely it is used,
compared with some measure of the potential user base.
Encouraging a target demographic
to download, install, start up and make meaningful use of a software
application is a difficult job, and providing an excellent product
is unfortunately no guarantee of this kind of success.

Human-Computer Interaction in this sense begins not when the
user is sat in front of a given running application,
but when she is considering how to get a certain job done.
Learning a new tool is {\em prima facie\/} a less attractive option
than making use of a familiar one,
so astronomers, like other humans, are generally resistant to
using new software.
With this in mind, it is important to make entry barriers low
and first impressions good:
installation and startup should be made as simple as possible,
and beginning use should provide rewarding results
in little time and with minimal effort.
Tutorials or classes
(a captive audience forced to get over the hurdle of initial use) 
can work well, as long as they leave a good impression.
Word of mouth is very useful, particularly an enthusiastic user
in a research group.
Tutorial and reference documentation and on-line teaching materials
ought to be provided,
though it is probably true that most users don't read them.
However, there is no magic recipe for this.
Many factors such as geographical and political ones are likely to
be beyond the control of the software developers,
and in any case encouraging take-up is a long job.
At the time of writing,
TOPCAT has been available for almost 15 years,
and citations are still rising more or less linearly,
which suggests that saturation of the potential user base
has yet to be reached.

\subsection{Defining Requirements}

A well-known, though possibly apocryphal, quote attributed to Henry Ford
claims that if he had asked his customers what they wanted, they would
have demanded faster horses.
Asking users what they want from software sounds like a promising idea,
but users in fact do not know.
Specifying requirements for data analysis software is a difficult job,
and astronomers are in general much too busy doing astronomy to
expend the necessary effort.
This task is really the responsibility of the software developer or
project team.
Top-down visionary design also sounds promising, but at least in
TOPCAT's case, the author also lacks the necessary abilities.
Instead, a policy is adopted of
incremental development informed by user engagement,
with a short iteration cycle.
Some functionality is tentatively implemented and published to users,
and feedback such as requests for additional related functionality
is used to direct future development.
New requirements can also be gathered from more general feedback.
Sometimes users will request or suggest a new feature which can
be implemented directly, but more often such input serves to indicate
the kind of problems that users are trying to solve, which can
inform the design of new, perhaps more general, capabilities.

\subsection{User Engagement}

Effective user engagement is an essential part of the requirement
gathering process.
All user engagement is considered beneficial,
and it is encouraged wherever possible.

Bug reports are particularly useful, indicating that a user is not only
using a particular feature, but cares enough about its correct
operation to file a report.
As well as identifying errors to be fixed,
they also provide a sample of
actual usage patterns that may suggest missing functionality
or opportunities for improvement.

A public user mailing list is provided for discussion of TOPCAT.
At time of writing it has around 150 subscribers and a few threads
per month.
In most cases queries are posted by users and answered quite promptly
by the author, though sometimes subscribers answer each other's questions.
In either case, replies offer an opportunity to publicise functionality that 
other list subscribers may not be aware of.
Use of the mailing list is not however obligatory for support requests;
users are always welcome to mail the author directly,
if they prefer private communication.

Preparing and delivering software tutorials and demonstrations
is difficult and time-consuming to do well,
but it is a valuable activity.
Apart from the (hopefully) positive promotional benefits,
constructing worked examples that use the software in question
to perform some useful scientific task can provide
a user's-eye view of missing, inadequate or broken functionality.
A developer required to demonstrate the software in action
in front of a live audience also has an excellent motivation to ensure
that the user interface and implementation are fit for purpose.

Contact with multiple projects is also helpful.
TOPCAT has been embedded in or funded under the auspices of
numerous different organisations,
from Starlink and AstroGrid in the early days
to various Euro-VO projects, GAVO, ESA, DPAC and others.
This somewhat nomadic existence has largely been 
driven by availability of funding,
but involvement with different software or astronomy projects
provides the benefit of exposure to different sets of
users, requirements or data holdings,
and this broadened perspective has helped to maintain the
appeal of the software to a wide user base.

\subsection{Prioritising Implementation}

Having gathered requirements for new or improved capabilities,
it is necessary to decide which items to implement from a
list which generally represents more than the available developer effort.
A useful rule of thumb is to do easy things first:
implementation effort is often not closely correlated with user benefit,
so that impact can often be improved by prioritising short jobs.

Beyond that, the main consideration when deciding what
capabilities to add to TOPCAT relates to the user interface.
It is all too easy to introduce a feature which only the application
author and the user who requested it know is present,
and such unknown features rarely represent efficient use of developer time.
New features must therefore be {\em discoverable\/},
and should preferably not degrade the user interface by
making existing features harder to use or to discover.
Demo-ware, expert-only controls, functionality which is highly data-specific,
and feature creep are all temptations to be resisted.
This is a difficult balancing act,
since adding new controls is bound to crowd out existing ones to some extent.
Managing it requires simultaneous attention to the constraints of
design, implementation and user support.

\subsection{GUI Design}

TOPCAT is a complex application with many functions.
Providing a Graphical User Interface that exposes all this
functionality in a way which is which is comprehensible and usable,
and preferably intuitive and unobtrusive, is perhaps the most
challenging problem its implementation presents.
A good GUI should be simple, flexible, responsive,
represent status visually,
indicate control functions using recognisable visual idioms,
and preclude or at least report erroneous control combinations.
Obstacles to this include limited availability of screen real
estate to accommodate controls and indicators of many-element
status, provision of recognisable idioms for novel functions,
and responsiveness for large datasets.
These problems become harder the more functionality is added,
and are especially acute for TOPCAT's visualisation windows,
which offer a very wide range of configuration options for
specifying plots.
GUI design in TOPCAT is far from a solved problem,
and it is likely that in many cases users are unaware of the
full range of functionality from which they could benefit.
Nevertheless we present some principles on which the GUI 
design has been based.

First, require minimal user effort.  The GUI should
do its best to leave the user's cognitive abilities free
for thinking about the science rather than about the software.
Where possible, the software should anticipate the user's
wishes to generate the desired result without any user
action at all, though the user must be free to override
such automatic decisions.
This can be achieved by ensuring that all controls take
suitable defaults.
Where explicit configuration actions are required,
it should be obvious how to take such actions,
for instance by selecting options from a list.
Providing an empty field for the user to fill in with some
numeric or other value is avoided if at all possible.

Secondly, the interface should be explorable.
The most important or commonly-required controls should
be placed in an obvious part of the screen, while less
essential controls should be reachable by exploration,
for instance by clicking on tabs with appropriate names
or buttons bearing suggestive images.
Finally, the effect of adjusting any control should be
instantly reflected in the display.
Once the user is familiar with basic operation, which should
be easy to master, she can experiment with other parts of
the GUI by viewing unfamiliar components and playing with
controls to see the effect they have on the display.
This facilitates a hands-on route to self-education about
the available capabilities and how to activate them.

These principles have been applied to TOPCAT's visualisation windows,
the most complex parts of its GUI.
When a 2-d plot window is opened, a plot is immediately displayed.
The plotted quantities are chosen automatically (the first two numeric
columns of the current table are plotted against each other),
the axes are automatically scaled to the data range,
and the {\em shading mode\/} is set to one that makes sense for
small or large datasets --- in sparse regions it resembles a scatter plot
and in dense regions it resembles a density map \citep{2014ASPC..485..257T}.
All configuration options have default values that combine to give a
reasonable-looking plot, and the same rule is followed when overlaying
other plot types such as contours or histograms.
Without making any decisions therefore,
the user is presented with a plot rather than a blank screen.
This is almost certainly not the plot the user wants to see,
but it provides a comprehensible
starting point from which various options can be
adjusted, according to the user's requirements
and the level of effort they are willing to expend.
Any change to the controls triggers an immediate replot
so the effect of any GUI interaction is instantly visible;
implementing this in a responsive manner entails considerable effort
\citep{informatics4030018},
but provides a much better user experience than requiring,
for instance, the user to hit a replot button.

\section{Conclusions}

This paper reviews the capabilities of the TOPCAT application
and presents some solved and unsolved problems encountered
during its development.
Implementation approaches are discussed for
a client-side analysis application capable of
retrieving quite large datasets from external services
and manipulating them locally.
Some other questions for which solutions are more elusive are
also discussed, such as requirement gathering, user engagement,
and GUI design.
These insights have been acquired during the development of
a particular software suite, but it is hoped that they may
be of interest to individuals or groups developing other
user software with similar aims.

\acknowledgements Development of the TOPCAT and STILTS software has been supported by many organisations over the years, including the UK's STFC and previously PPARC research councils, EU FP6 and FP7 programmes, GAVO and ESA.

\bibliography{I3-1}  

\end{document}